# Magnetization reversal of thin ferromagnetic elements with surface anisotropy

N. A. Usov[1,2] and O. N. Serebryakova[1,2]

[1]*National University of Science and Technology «MISIS», 119049, Moscow, Russia*
[2]*Pushkov Institute of Terrestrial Magnetism, Ionosphere and Radio Wave Propagation, Russian Academy of Sciences, (IZMIRAN) 108480, Troitsk, Moscow, Russia*

**Abstract.** The magnetization reversal process in thin-film ferromagnetic elements with surface anisotropy of various shapes and sizes is investigated by means of numerical simulation. The dependence of the perpendicular and in-plane hysteresis loops on the element thickness and the value of the surface anisotropy constant is obtained. For sufficiently large values of the surface anisotropy constant the magnetization reversal of thin-film elements is shown to occur due to the nucleation of the buckling mode. For an elongated rectangular element the nucleation field of the buckling mode is proportional to the absolute value of the surface anisotropy constant, and inversely proportional to the element thickness.

## Introduction

The properties of thin ferromagnetic films with surface anisotropy are promising for applications of such magnetic materials in modern thin-film electronics devices [1-4]. From phenomenological point of view, the effect of surface anisotropy on the magnetization distribution in a ferromagnet can be described by introducing a special energy contribution to the total sample energy that depends on the orientation of the unit magnetization vector $\alpha(r)$ at the sample surface. Taking into account symmetry considerations, the energy density of the surface magnetic anisotropy in a simplest case can be written in the form [5-8]

$$w_{sa} = K_s (\vec{\alpha}\vec{n})^2, \qquad (1)$$

where $K_s$ is the surface anisotropy constant, $n$ being the unit vector perpendicular to the sample surface. If surface magnetic anisotropy constant $K_s$ is negative and large enough in absolute value, a reorientation of the unit magnetization vector perpendicular to the film surface is energetically favorable, in spite of the increase in the magnetostatic energy of the ferromagnetic sample. The reorientation of the unit magnetization vector with a change in the ferromagnetic film thickness was observed in the number of experiments with thin films of iron, cobalt, and other ferromagnets [9-12].

Currently, it is supposed that the surface magnetic anisotropy can have various origins. It can be related with the specifics of the spin-orbit interaction on the sample surface [13], with the difference in the atomic lattices periods of the sample and substrate [14], with the distribution of inhomogeneous mechanical stresses near the interface [15], etc. However, from a point of view of the Micromagnetics [5] it is important that the energy contribution, Eq. (1), is concentrated in a very narrow region near the sample surface. Thus, it is a surface contribution to the total sample energy. As a result, its effect on the distribution of the unit magnetization vector in the volume of the ferromagnet can be properly described by the corresponding boundary condition acting on the sample surface.

This general approach has recently been consistently carried out [16] to study the equilibrium magnetization distributions in a thin ferromagnetic film with surface anisotropy by means of numerical simulation. It was shown [16] that if the surface anisotropy constant $K_s$ is less than a certain critical value, then there exists the so-called spin canted micromagnetic state in the intermediate range of film thicknesses, $L_{z,min} < L_z < L_{z,max}$, with average magnetization inclined at some angle to the sample surface. For sufficiently small thicknesses, $L_z < L_{z,min}$, the film is magnetized perpendicular to the surface (z-state), while for $L_z > L_{z,max}$, the magnetization lies in the film plane. On the other hand, if the surface anisotropy constant exceeds the critical value, different labyrinth domain structures are realized [16] near the surface of the film of sufficiently large thickness.

In this paper, the magnetization reversal process in thin-film ferromagnetic elements with surface anisotropy is studied by means of numerical simulation. Both in-plane and out of plane quasi-static hysteresis loops of thin film elements of various shapes, thickness and in-plane dimensions are calculated. By comparison with the experimental data, the numerical results would allow one to determine the value of the surface anisotropy constant of thin ferromagnetic film studied.

## Numerical simulation

The numerical simulation is carried out for thin ferromagnetic nano elements of various thicknesses, $L_z = $ 4-6 nm. We consider elements of circular and rectangular shape with different in-plane aspect ratios, $L_x/L_y = 1.0 - 3.0$, the in-plane dimensions of the elements being of the order of 100 - 400 nm. The elements are assumed to be cut out of a thin amorphous ferromagnetic CoSiB film [1, 2], with surface anisotropy, and with induced magnetic anisotropy in the plane of the film. The saturation magnetization of the CoSiB film is given by $M_s = 500$ emu/cm$^3$, the exchange constant $C = 2\times10^{-6}$ erg/cm, the anisotropy constant of the induced volume anisotropy is assumed to be $K_V = 10^4$ erg/cm$^3$. In the calculations performed the surface anisotropy constant varies within the range $K_s = -(0.6 - 1.2)$ erg/cm$^2$.

For simplicity, it is further assumed that the surface anisotropy is present only at the upper boundary of the film. In this case, the micromagnetic boundary condition for the unit magnetization vector at $z = L_z$ has the form [5]

$$C\frac{\partial \alpha_x}{\partial z} = -2|K_s|\alpha_z^2\alpha_x; \quad C\frac{\partial \alpha_y}{\partial z} = -2|K_s|\alpha_z^2\alpha_y;$$
$$C\frac{\partial \alpha_z}{\partial z} = -2|K_s|\alpha_z(\alpha_z^2 - 1). \qquad (2)$$

The usual boundary condition, $\partial \vec{\alpha}/\partial n = 0$, acts on other surfaces of the film.

The magnetization reversal process in thin-film elements with surface anisotropy is studied by solving the Landau-Lifshitz-Gilbert equation

$$\frac{\partial \vec{\alpha}}{\partial t} = -\gamma[\vec{\alpha}, \vec{H}_{ef}] + \kappa\left[\vec{\alpha}, \frac{\partial \vec{\alpha}}{\partial t}\right], \qquad (3)$$

where $\gamma$ is the gyromagnetic ratio, and $\kappa$ is the phenomenological magnetic damping constant. The total effective magnetic field in the volume of the film takes into account the exchange, anisotropic, and magneto-dipole interactions

$$\vec{H}_{ef} = \frac{C}{M_s}\Delta\vec{\alpha} + \vec{H}_0 + \vec{H}' - \frac{\partial w_a}{M_s\partial\vec{\alpha}}. \qquad (4)$$

Here $\vec{H}'$ is the demagnetizing field, $\vec{H}_0$ is the external applied magnetic field, $w_a = K_V(\alpha_y^2 + \alpha_z^2)$ is the energy density of the induced magnetic anisotropy with the easy anisotropy axis lying in the plane of the film and oriented along the $x$ axis.

For numerical simulation the ferromagnetic film is approximated by small cubic numerical cells with an edge $b = 1.5-3$ nm sufficiently small in comparison with the exchange length of the ferromagnet, $L_{ex} = \sqrt{C}/M_s \approx 28$ nm. In the calculations performed the thin film elements were approximated by a sufficiently large number of numerical cells, $N \sim 10^3 - 10^4$, in order to maintain an accuracy of the numerical results obtained.

The calculation of the quasi-static hysteresis loop of a thin film element begins with finding a quasi-homogeneous micromagnetic configuration in a sufficiently strong external magnetic field applied in-plane, or perpendicular to the film plane, respectively. Then, the external magnetic field decreases by a small value, $dH_0 = 1 - 2$ Oe, and the evolution of the unit magnetization vector $\alpha(r)$ is calculated in accordance with Eqs. (2) - (4), until a new equilibrium micromagnetic state is reached. The magnetization distribution in applied magnetic field is considered to be stable when the criterion

$$\max_{(1\leq i\leq N)}\left\|[\vec{\alpha}_i, \vec{H}_{ef,i}/\|\vec{H}_{ef,i}\|]\right\| < 10^{-6} \qquad (5)$$

is fulfilled. This means that the maximum deviation of the unit magnetization vector from the direction of the effective magnetic field in the same numerical cell does not exceed a small predefined value.

**Results and discussion**

**In-plane hysteresis loops**

Fig. 1 shows the in-plane quasi-static hysteresis loops of rectangular thin film elements with various aspect ratios calculated numerically for different values of the surface anisotropy constant. One can see that the remanent magnetization of the elements decreases with increasing of the absolute value of the surface anisotropy constant. Moreover, the larger the absolute value of this constant, the greater the magnetic saturation field of the element. In addition, with an increase in $|K_s|$ the area of the hysteresis loop decreases substantially. The existence of in-plane remanent magnetization corresponds to the spin canted magnetization state. The remanent magnetization is zero for perpendicular magnetized $z$-state.

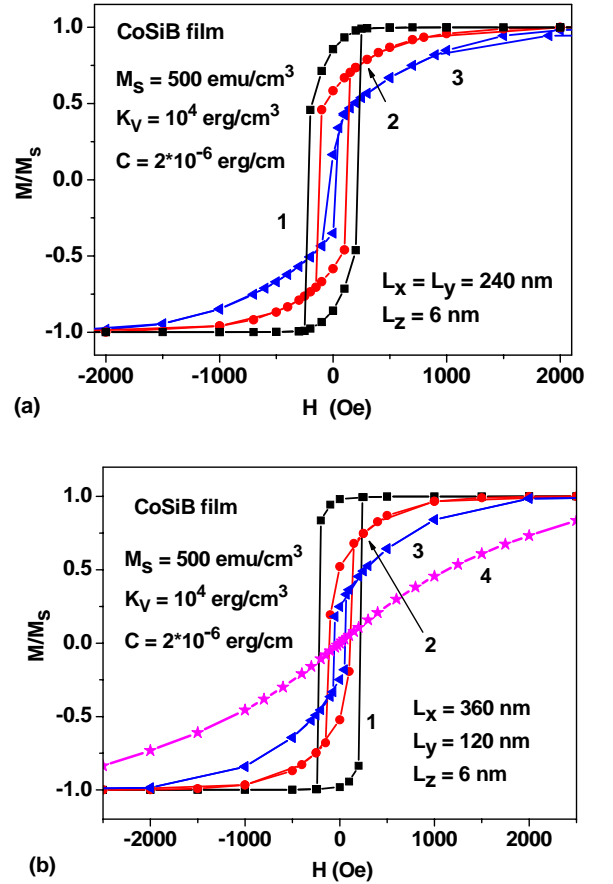

Fig. 1 Quasistatic hysteresis loops of CoSiB elements with different aspect ratios: a) $L_x/L_y = 1.0$; b) $L_x/L_y = 3.0$. The magnetic field is applied in the plane of the elements along the $x$ axis. The calculations are carried out for different values of the surface anisotropy constant: $K_s = -0.7$ erg/cm$^2$; 2) $K_s = -0.9$ erg/cm$^2$; 3) $K_s = -1.0$ erg/cm$^2$; 4) $K_s = -1.25$ erg/cm$^2$.

This case is exemplified by the curve 4) in Fig. 1b. In principle, such behavior of the hysteresis loops is expected, since the presence of surface anisotropy with $K_s < 0$ promotes the deviation of the unit magnetization vector perpendicular to the film plane, while the external magnetic field tries to orient the magnetization vector along the film surface.

In general, the calculation of an in-plane hysteresis loop for a thin film element with surface anisotropy can be performed only by numerical simulation. At the same time, in the case of a perpendicularly magnetized $z$-state the magnetization curve of a thin ferromagnetic film in a magnetic field $H_0$ applied along the film surface can be calculated analytically. Assuming the uniform rotation of the unit magnetization vector in this case, the total energy of the film per unit area can be written as [16]

$$w = \left(K_{ef}L_z - |K_s|\right)\cos^2\theta - M_s H_0 L_z \sin\theta, \quad (6)$$

where $K_{ef} = K_V + 2\pi M_s^2$, $\theta$ is the angle of the unit magnetization vector with respect to the $z$ axis.

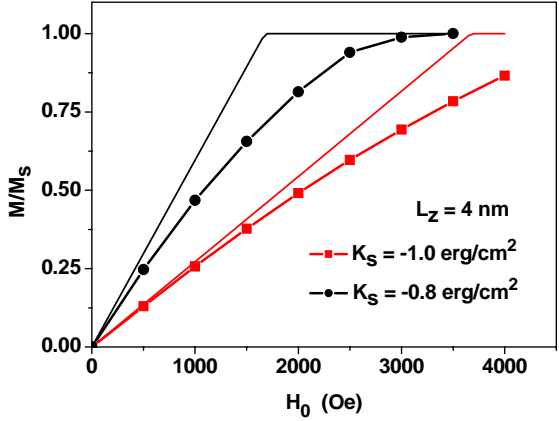

Fig. 2. Magnetization curves (dots) of the perpendicular magnetized $z$-state of rectangular element with thickness $L_z = 4$ nm and in-plane sizes $L_x = 240$ nm, $L_y = 80$ nm as the function of in-plane magnetic field for various values of the surface anisotropy constant. The solid lines are drawn in accordance with Eq. (7).

Minimizing Eq. (7) as a function of the angle $\theta$, one obtains the equilibrium value of $x$-component of the unit magnetization vector as a function of the applied magnetic field $H_0$

$$\alpha_x = \sin\theta = \begin{cases} H_0/H_s, & H_0 \leq H_s \\ 1, & H_0 > H_s \end{cases}, \quad (7)$$

where $H_s = 2(|K_s|/L_z - K_{ef})/M_s$ is the corresponding saturation field of the film.

As an example, Fig. 2 shows the magnetization curves of perpendicular magnetized $z$-state in a thin-film element with thickness $L_z = 4$ nm calculated numerically for different values of the surface anisotropy constant. The solid curves in this figure are drawn using Eq. (7). Some deviation of numerical results from Eq. (7) is explained by the influence of the demagnetizing field of the element of finite in-plane size. The in-plane demagnetizing field is created by magnetic charges that arise at the side ends of the element under the influence of in-plane magnetic field. This effect is not taken into account in Eqs. (6) and (7), which are valid for an infinite film. At the same time, as can be seen from Fig. 2, the slope of the magnetization curves in small magnetic fields is in satisfactory agreement with Eq. (7).

**Perpendicular hysteresis loops**

Let us now consider the magnetization reversal process of a thin film element in magnetic field perpendicular to the film plane. In this case, the study of the magnetization reversal of the $z$- state is most interesting, since in this case the numerical results can be compared with the analytical theory of the nucleation fields [5]. It is well known [5], that the magnetization reversal of a homogeneously magnetized ferromagnetic sample is initiated in a reversed applied magnetic field, which is called the nucleation field. The value of the nucleation field and the mode of the magnetization reversal can depend on various factors, in particular, on the shape, size and in-plane aspect ratio $L_x/L_y$ of rectangular element.

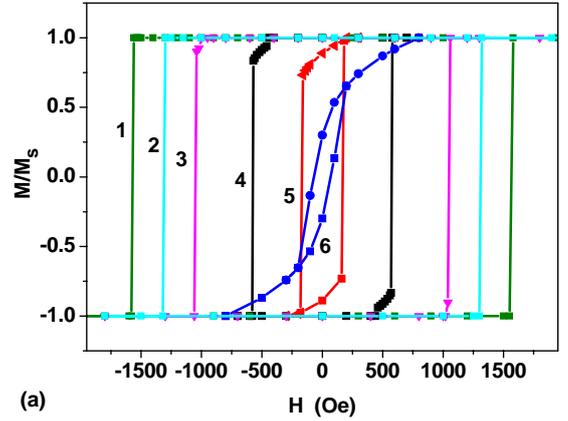

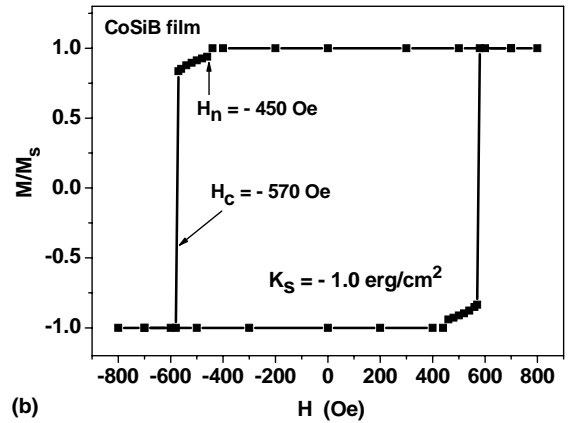

Fig. 3. a) Perpendicular hysteresis loops of rectangular CoSiB element with dimensions $L_x = 360$ nm, $L_y = 120$ nm, $L_z = 6$ nm, depending on the value of the surface anisotropy constant: 1) $K_s = -1.2$ erg/cm$^2$; 2) $K_s = -1.15$ erg/cm$^2$; 3) $K_s = -1.1$ erg/cm$^2$; 4) $K_s = -1.0$ erg/cm$^2$; 5) $K_s = -0.9$ erg/cm$^2$; 6) $K_s = -0.8$ erg/cm$^2$; b) Nucleation field and coercive force of the element with $K_s = -1.0$ erg/cm$^2$.

Fig. 3a shows the perpendicular hysteresis loops of rectangular thin-film element with aspect ratio $L_x/L_y = 3.0$, depending on the value of the surface anisotropy constant. One can see that the coercive force of the perpendicular hysteresis loops of this element increases with increasing of $|K_s|$. For sufficiently large absolute values of $|K_s| \geq 1.15$ erg/cm$^2$ the hysteresis loops 1) and 2) in Fig. 3a are strictly rectangular. This means that for $|K_s| \geq 1.15$ erg/cm$^2$ the magnetization reversal process occurs in a strong negative magnetic field in a single Barkhausen jump.

On the other hand, for $|K_s|$ in the interval $1.0 \leq |K_s| \leq 1.15$ erg/cm$^2$ (loops 3) and 4) in Fig. 3a), a nonlinear stabilization of the nucleation mode occurs in a certain range of magnetic fields just after the mode nucleation, where an intermediate non-uniform micromagnetic state exists. At the end of this interval the complete reversal of the element takes place in Barkhausen jump.

Finally, for the values $|K_s| \leq 0.9$ erg/cm$^2$, the magnetization of the element is nonuniform even in zero magnetic field. Actually, it follows from an earlier analysis [16] that with a decrease in $|K_s|$ the perpendicularly magnetized z-state of the element transforms into the spin canted state. The hysteresis loops 5), 6) in Fig. 3a correspond to the spin canted micromagnetic state, for which, in the absence of an external magnetic field, the average value of the unit magnetization vector is inclined at some angle to the surface of the element.

Fig. 3b shows in detail the magnetization reversal process of the same thin-film element with surface anisotropy constant $K_s = -1.0$ erg/cm$^2$. In this case the nucleation field at the element is given by $H_n = -450$ Oe, whereas the Barkhausen jump that determines the coercive force of the element occurs in the field $H_0 = -570$ Oe. Consequently, a stable inhomogeneous micromagnetic state in this element exists within the interval $-570 < H_0 < -450$ Oe.

Fig. 4 shows the perpendicular hysteresis loops of a circular thin film element with diameter $D = 180$ nm and thickness $L_z = 6$ nm, depending on the value of the surface anisotropy constant. One can see that the behavior of the perpendicular hysteresis loops is similar to that of Fig. 3, but the numerical values of the nucleation fields for circular element differ from that of the rectangular sample.

**Nucleation mode**

In general, the study of the nonlinear stabilization of the nucleation mode and the determination of the coercive force of a thin film element can be performed only by numerical simulation. As an example, Fig. 5 shows successive steps 1) - 3) of the magnetization reversal process for rectangular and circular thin film elements during the corresponding Barkhausen jumps. In Fig. 5 the areas where the $\alpha_z$ component of the unit magnetization vector is positive or negative are shown in red and in blue, respectively. Fig. 5a corresponds to the Barkhausen jump on curve 1) in Fig. 3a, whereas Fig. 5b show the evolution of the element magnetization during the Barkhausen jump on curve 3) in Fig. 4, respectively. As Fig. 5 shows, for both elements after the loss of stability of the homogeneous state, the magnetization reversal proceeds owing to the formation and rapid propagation of the domain walls along the sample.

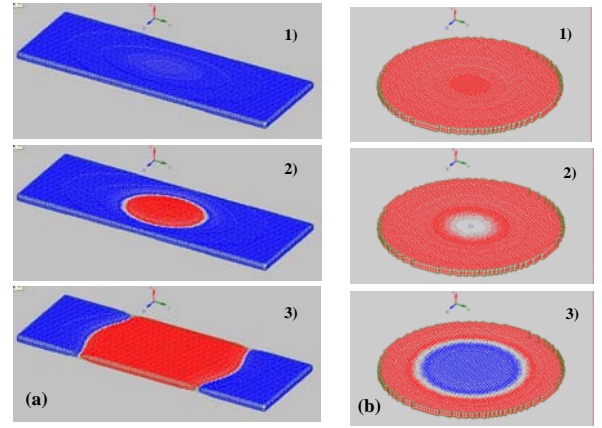

Fig. 5. Dynamics of magnetization reversal in thin film elements: a) rectangular element with dimensions $L_x = 360$ nm, $L_y = 120$ nm, $L_z = 6$ nm and surface anisotropy constant $K_s = -1.2$ erg/cm$^2$ in magnetic field $H_c = -1580$ Oe; b) circular element with diameter $D = 180$ nm and thickness $L_z = 6$ nm, with a surface anisotropy constant $K_s = -1.0$ erg/cm$^2$ in magnetic field $H_c = -700$ Oe.

It should be noted that Fig. 5 shows only the late, nonlinear stage of magnetization reversal of the elements during the irreversible Barkhausen jumps. The latter occurs in a magnetic field corresponding to the coercive force of the element $H_0 = H_c$. At the same time, when the instability mode is just nucleated in the nucleation field, $H_0 = H_n$, the amplitude of the mode is very small. Therefore the general nonlinear micromagnetic equations allow linearization [5]. The investigation of the nucleation fields of uniformly magnetized small ferromagnetic samples is one of the most important micromagnetic problems [5]. Unfortunately, the analytical investigation of the problem is possible [5] only for samples of high symmetry (sphere, long cylinder). Nevertheless, the value of the nucleation field and the shape of the corresponding lowest nucleation mode of the ferromagnetic element in the reversed magnetic field can be determined by means of the numerical simulation.

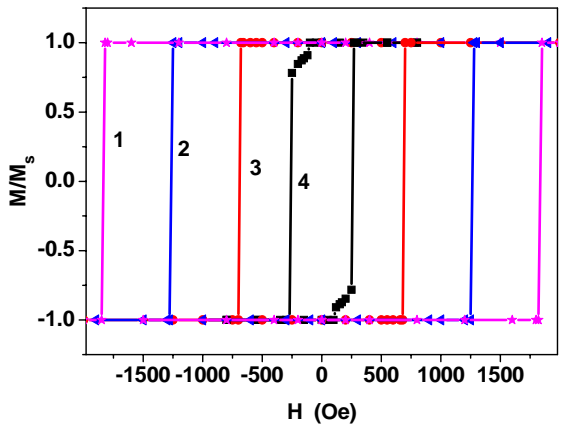

Fig. 4. Perpendicular hysteresis loops of a circular CoSiB element with diameter $D = 180$ nm and thickness $L_z = 6$ nm depending on the value of the surface anisotropy constant: 1) $K_s = -1.2$ erg/cm$^2$; 2) $K_s = -1.1$ erg/cm$^2$; 3) $K_s = -1.0$ erg/cm$^2$; 4) $K_s = -0.9$ erg/cm$^2$.

Fig. 6a shows the shape of the nucleation mode for the rectangular element with dimensions $L_x$ = 360 nm, $L_y$ = 120 nm, $L_z$ = 6 nm and surface anisotropy constant $K_s$ = -1.2 erg/cm$^2$ at the initial stage of the magnetization reversal in the magnetic field $H_c$ = -1580 Oe. The magnetization distribution shown in Fig. 6a corresponds to the well known buckling mode, which exists in small elongated ferromagnetic samples [5]. For rectangular element with the aspect ratio $L_x/L_y$ = 3.0, the shape of the nucleation mode can be approximated with a good accuracy by the equations

$$\alpha_x = A\sin(\pi x/L_x); \qquad \alpha_y = 0.0;$$
$$\alpha_z = -\sqrt{1-\alpha_x^2 - \alpha_y^2}, \qquad (8)$$

where $A$ is the mode amplitude, small at the initial stage of the reversal. Fig. 6b, 6c show the nonlinear evolution of the buckling mode at the late stages of the development of this instability (see also Fig. 5a).

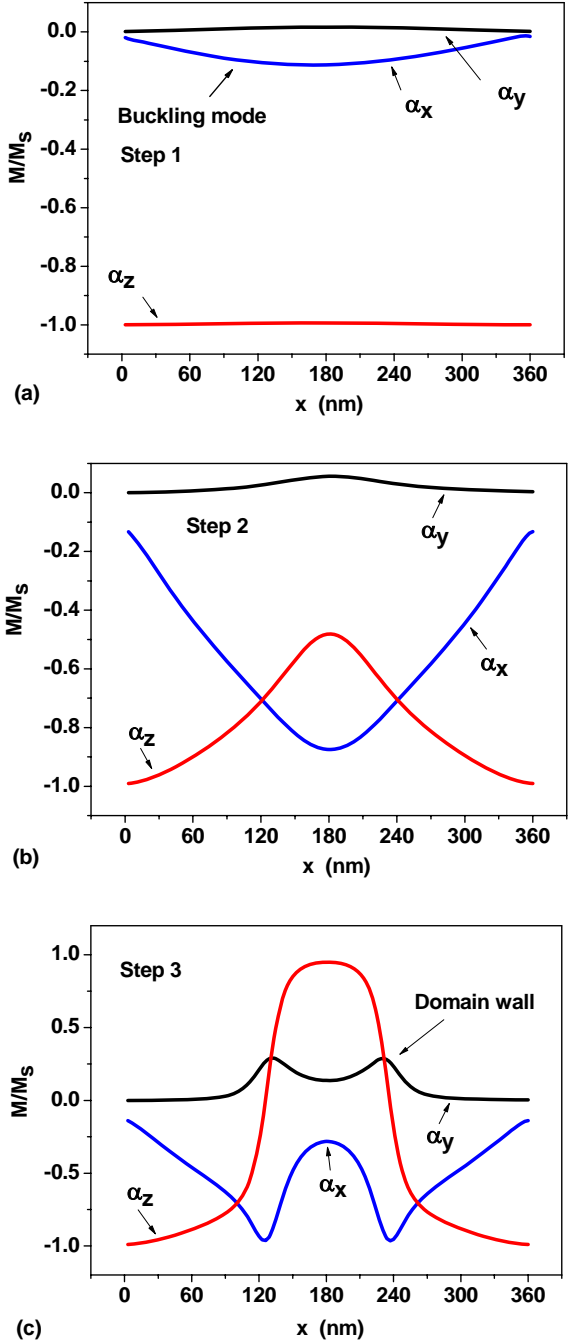

Fig. 6. The evolution of the magnetization distribution in the rectangular element with dimensions $L_x$ = 360 nm, $L_y$ = 120 nm, $L_z$ = 6 nm and surface anisotropy constant $K_s$ = -1.2 erg/cm$^2$ along the line $y = L_y/2$, $z = L_z$ during magnetization reversal in applied magnetic field $H_c$ = -1580 Oe: a) the shape of the buckling nucleation mode at the initial stage of the process, b), c) the transient inhomogeneous magnetization distributions at the subsequent stages of the process.

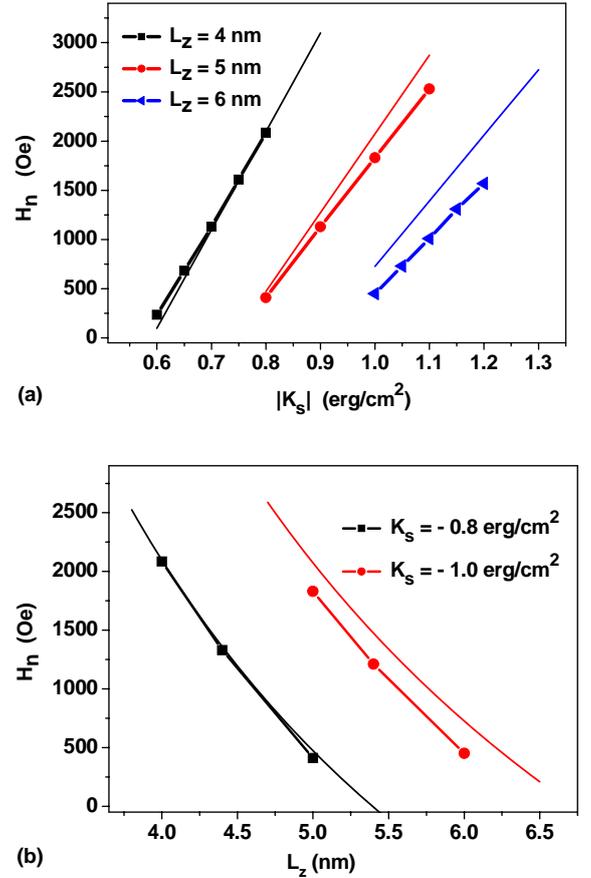

Fig. 7. The nucleation field of the buckling mode in rectangular samples with the aspect ratio $L_x/L_y$ = 3.0: a) as a function of $|K_s|$ for different film thickness $L_z$; b) as a function of $L_z$ for fixed values of the surface anisotropy constant. The solid lines are drawn according to Eq. (12).

Figs. 7a, 7b show calculated numerically (dots) the dependence of the nucleation field of rectangular thin-film samples with aspect ratio $L_x/L_y$ = 3.0 on the value of the surface anisotropy constant, and on the thickness of the element, respectively. For thickness $L_z$ = 6 nm the in-plane dimensions of the element are given by $L_x$ = 360 nm, $L_y$ = 120 nm, for smaller thickness the in-plane dimensions are scaled accordingly. Since the shape of the nucleation mode for this element is known (see Eq. (8)),

the value of the corresponding nucleation field can be approximately determined by means of a variational estimate [5].

The second order correction to the total energy of the z-state of the rectangular element due to the magnetization perturbation of the first, $\vec{\alpha}^{(1)} = (\alpha_x, 0, 0)$, and the second, $\vec{\alpha}^{(2)} = (0, 0, -\alpha_x^2/2)$, orders of magnitude has the form

$$\delta W^{(2)} = \frac{C}{2}\int dv (\vec{\nabla}\alpha_x)^2 + |K_s|\int ds\, \alpha_x^2 - \left(\frac{M_s(H_0 + |H_z'^{(0)}|)}{2} + K_V\right)\int dv\, \alpha_x^2 + \frac{1}{8\pi}\int dv(\vec{H}'^{(1)})^2 \quad (9)$$

Here the first term gives the perturbation of the exchange energy of the element, the second term, where the integral is taken over the element surface, $z = L_z$, gives the perturbation of the surface anisotropy energy, the third term is the Zeeman energy perturbation, the perturbation of the volume magnetic anisotropy, and part of the magnetostatic energy perturbation. The unperturbed demagnetizing field of the element can be expressed through the demagnetizing factor of the sample, $H_z'^{(0)} = -N_z M_s$. Finally, the last term in Eq. (9) is the magnetostatic energy associated with the first-order magnetization perturbation, $\vec{\alpha}^{(1)}$. Due to the small thickness of the element, the main contribution to this energy is the magnetostatic energy of the volume magnetic charges $\rho(x)$,

$$\delta W_m^{(2)} = \frac{1}{8\pi}\int dv(\vec{H}'^{(1)})^2 = \frac{1}{2}\int dv\, dv_1 \frac{\rho(x)\rho(x_1)}{|\vec{r}-\vec{r}_1|}, \quad (10a)$$

distributed in the volume of the element with the density

$$\rho(x) = -M_s \frac{d\alpha_x}{dx} = -AM_s \frac{\pi}{L_x}\cos\left(\frac{\pi x}{L_x}\right). \quad (10b)$$

Taking into account the small thickness of the element, the magnetostatic energy, Eq. (10), can be evaluated as follows

$$\delta W_m^{(2)} = \frac{(\pi M_s A)^2}{2} V \frac{L_z}{L_x} Q\left(\frac{L_x}{L_y}\right), \quad (11)$$

where $V = L_x L_y L_z$ is the volume of the element, the dimensionless function $Q(\xi)$ being

$$Q(\xi) = \int_0^1 dx \int_0^1 dx_1 \int_0^1 dy\, \cos(\pi x)\cos(\pi x_1)$$
$$\ln\frac{1 + y + \sqrt{\xi^2(x-x_1)^2 + (1+y)^2}}{y + \sqrt{\xi^2(x-x_1)^2 + y^2}}.$$

The maximum value of this function calculated numerically is given by $Q_{max} \approx 0.12$ at $\xi = 5.4$. Then it decreases as the function of parameter $\xi$, $Q \sim 1/\xi$.

Equating the total energy correction, Eq. (9), to zero, one obtains the nucleation field of the buckling mode of the rectangular thin film element

$$H_n = \frac{C}{M_s}\left(\frac{\pi}{L_x}\right)^2 + \frac{2|K_s|}{M_s L_z} - |H_z'^{(0)}| - \quad (12)$$
$$\frac{2K_V}{M_s} + 2\pi^2 M_s \frac{L_z}{L_x} Q\left(\frac{L_x}{L_y}\right).$$

For sufficiently large in-plane dimensions of the element, $L_x$, $L_y \gg L_z$, the main contributions to this expression are given by the second term, which describes the effect of surface anisotropy, and the third term, which gives a large demagnetizing field of the element magnetized perpendicular to the plane, $|H_z'^{(0)}| \approx 4\pi M_s$. For example, for the thin film element with dimensions $L_x$ = 240 nm, $L_y$ = 80 nm, $L_z$ = 4 nm and surface anisotropy constant $K_s$ = -0.6 erg/cm$^2$ the values of the successive terms in Eq. (12) are given by 68.5, 6000, -5950, -40 and 19.5 Oe, respectively.

Since the demagnetizing field of a thin film element only slightly changes as the function of its thickness in the range of small thicknesses $L_z$ = 4 - 6 nm, the nucleation field of the buckling mode turns out to be proportional to the absolute value of the surface anisotropy constant $|K_s|$, and inversely proportional to the thickness of the element $L_z$. The solid lines in Fig. 7 are drawn in accordance with Eq. (12). As Fig. 7 shows, the variation estimate of the nucleation field, Eq. (12), is in a qualitative agreement with the numerical calculations of the nucleation fields of rectangular thin-film elements with surface anisotropy.

**Conclusions**

In this paper, the magnetization reversal process in thin film ferromagnetic elements with surface anisotropy of various shapes and sizes is investigated. The dependence of the perpendicular and in-plane hysteresis loops on the element thickness, and the value of the surface anisotropy constant is studied. It is shown that for sufficiently large values of the surface anisotropy constant, the magnetization reversal of elongated thin film elements is due to nucleation of the buckling mode. For the rectangular element the approximate formula is obtained for the dependence of the nucleation field of the buckling mode on the element thickness and the value of the surface anisotropy constant.

The investigation of the magnetization reversal process in thin-film elements with surface anisotropy seem to be useful for further improvement of the technology of preparation of thin-film elements with the magnetic characteristics necessary for various applications. It seems to be helpful also for the accurate determination of the surface anisotropy constant $K_s$ from the experiment.

**Acknowledgement**

The authors wish to acknowledge the financial support of the Ministry of Education and Science of the Russian Federation in the framework of Increase Competitiveness Program of NUST «MISIS», contract № K2-2015-018.

**References**

[1] Q.L. Ma, S. Iihama, T. Kubota, X.M. Zhang, S. Mizukami, Y. Ando, and T. Miyazaki, Appl. Phys. Lett. **101,** 122414 (2012).

[2] J. Yoon, S. Jung, Y. Choi, J. Cho, C.-Y. You, M.H. Jung, and H.I. Yim, J. Appl. Phys. **113**, 17A342 (2013).

[3] S. Mangin, D. Ravelosona, J.A. Katine, M.J. Carey, B.D. Terris, and E.E. Fullerton, Nature Mater. **5,** 210 (2006).

[4] C.-H. Lambert, A. Rajanikanth, T. Hauet, S. Mangin, E.E. Fullerton, and S. Andrieu, Appl. Phys. Lett. **102,** 122410 (2013).

[5] W.F. Brown, Jr., *Micromagnetics* (Wiley-Interscience, New York - London, 1963).

[6] M.T. Johnson, P.J.H. Bloemen, F.J.A. den Broeder, J.J. de Vries, Rep. Prog. Phys. 59 (1996) 1409.

[7] H.N. Bertram, D.I. Paul, J. Appl. Phys. 82 (1997) 2439.

[8] C.A.F. Vaz, J.A.C. Bland, G. Lauhoff, Rep. Prog. Phys. **71**, 056501 (2008).

[9] R. Allenspach, M. Stampanoni, and A. Bischof, Phys. Rev. Lett. **65**, 3344 (1990).

[10] H. Fritzsche, J. Kohlepp, H.J. Elmers, and U. Gradmann, Phys. Rev. B **49**, 15665 (1994).

[11] M. Speckmann, H. P. Oepen, and H. Ibach, Phys. Rev. Lett. **75**, 2035 (1995).

[12] M. Hehn, S. Padovani, K. Ounadjela, and J. P. Bucher, Phys. Rev. B **54**, 3428 (1996).

[13] H.X. Yang, M. Chshiev, B. Dieny, J.H. Lee, A. Manchon, and K.H. Shin, Phys. Rev. B **84,** 054401 (2011).

[14] K.H. He and S.J. Chen, J. Appl. Phys. **111,** 07C109 (2012).

[15] J.I. Hong, S. Sankar, A.E. Berkowitz, and W.F. Egelhoff Jr., J. Magn. Magn. Mater. **285,** 359 (2005).

[16] N. A. Usov and O. N. Serebryakova, J. Appl. Phys. **121,** 133905 (2017).